\documentclass[12pt]{article}

\usepackage{sbc-template}
\usepackage{graphicx,url}
\usepackage[utf8]{inputenc}
\usepackage{listings}

\title{An agent-based approach to procedural city generation incorporating Land Use and Transport Interaction models}

\author{Luiz F. S. Eugênio dos Santos\inst{1}, Claus Aranha\inst{2}, André P. de L. F. de Carvalho\inst{1} }

\address{Instituto de Ciências Matemáticas e de Computação - University os São Paulo (USP)\\
São Carlos, São Paulo - Brazil
\nextinstitute
University of Tsukuba\\
  Tsukuba, Ibaraki - Japan
}

\begin{document} 

\maketitle

\begin{abstract}
    We apply the knowledge of urban settings established with the study of Land Use and Transport Interaction (LUTI) models to develop reward functions for an agent-based system capable of planning realistic artificial cities. The system aims to replicate in the micro scale the main components of real settlements, such as zoning and accessibility in a road network. Moreover, we propose a novel representation for the agent's environment that efficiently combines the road graph with a discrete model for the land. Our system starts from an empty map consisting only of the road network graph, and the agent incrementally expands it by building new sites while distinguishing land uses between residential, commercial, industrial, and recreational.
\end{abstract}

\section{Introduction}

\textit{Land Use and Transport Interaction} (\textit{LUTI}) models seek to address the problem of modeling cities based on the behavior and relation of the entities that compose them. The demand for such a model comes from the necessity by cities' residents to move spatially to perform different actions in separated spaces. This flow of people and vehicles, and the paths of commuters is of interest to economists and urban planners. The organization and behavior of the population in cities is thus the basis from which we explain and predict the appreciation or depreciation of regions, market prospecting and even many urban problems, such that understanding the parts that make up these intricate systems is of great importance.

The study of LUTI models is done through the analysis of how residents organize themselves within the city. That is, where they live, where they work, where commercial and leisure establishments are available, as well as the travel costs between regions, defined as the distance (considering the road network), the travel time, and the access to transport \cite{book1}.

With the increase in computation capacity and the demand for more detailed models, along with the restrictions imposed by previous models in scenarios with lack of data, new approaches have been developed. Techniques based on cellular automata, dynamic systems and agents have gained space in this domain \cite{cellular, perez, multiagent}. Out of those, the last one is of specially interesting for its extensive ability to model the interaction of agents with each other as well as with their environment, while fulfilling objectives in a scenario as complex as one wants to model and is able simulate \cite{Bouanan:2018}.

On the other hand, the procedural generation of artificial cities is another domain that aims to achieve a realistic representation of the urban environment. Being used extensively in the entertainment industry and especially in games, different approaches to determine the geometry and organization of the generated cities vary greatly from one another, and especially due to the lack of objective metrics, the struggles in evaluating models and results are a known characteristic in this domain \cite{kelly, muller, proceduralluti2006}. Consequently, while recent works continue to aim for an approach that realistically reproduces the shape of a city, in some cases even allowing the depiction of complex road networks \cite{nishida}, little is considered when it comes to the usability of the resulting maps for real residents, thus limiting the use of these models on urban planning.

Seeking to adapt the use of known methods in procedural generation while enabling them this purpose, algorithms that use as a basis the knowledge of the urban scenario built through the \textit{LUTI} models have been showing significant results \cite{townsim, rui, Groenewegen2009ProceduralCL}. This is especially true for agent-based models \cite{townsim}, where both choosing environment characteristics and the agents' goals and actions are extremely flexible while also providing a greater level of interaction with the entities that make up the map, consequently producing more detailed and custom-tailored outcomes.

In this context, the present works aims to apply the urban modelling tools commonly used to establish the value of certain regions in the city in the elaboration of an agent-based system capable of generating realistic artificial cities. We do so by first considering the usability of our system with real world data, which is done by taking as input a road network graph extracted from \textit{OpenStreetMap}'s files. The environment along with the reward functions and states are then modelled after utility theory-based \textit{LUTI} models \cite{Santander} and with the support of a Deep Q-Network, the agent learns to populate the map with residential, commercial, industrial and recreational land uses, while taking into account the overall accessibility of the cells considering the network as well as how the general positions of the zones interfere with the values of one another.

From our results, we validate the possibility of efficiently generating the land use for individual land plots while considering both local information of the development types of the neighboring lots as well as global information of the whole map, represented by the accessibility scores of each location when travelling in the city using the road network.

\section{System description}

In this paper, we propose an agent-based system for the procedural generation of cities \footnote{https://github.com/LFRusso/autoplanner}. We start with a road network as input and agent's parameters such as its view radius and weights used when calculating the score of each development type. The agent then explores the map, moving to each available plot from most to least accessible and classifying them as commercial, residential, industrial or recreational sites.

The agent is able to collect information about their neighborhood within its given view radius, which it can use in the decision-making process to reach its goal of maximizing the reward given by the sum of the scores for all developments. The final state is reached when all sites are developed or when the predefined maximum number of iterations is reached.

One of the main challenges in modeling urban environments is the integration and representation of the dynamics between the road network and the land, which consists of buildings, facilities, parks and so on. This difficulty stems from the fact that the road network is commonly modeled as a directed weighted graph, with attributes such as length, speed limit, and intersections, while the land lots are represented as shapes and rely heavily in attributes such as their position in relation to each other and globally on the map. 

To tackle this problem, we establish a link between the land use grid, composed of cells each representing a square plot of land with size of $400$ square meters, and the road network, loaded directly from an OpenStreetMap (OSM) XML-formatted file or Eclipse SUMO \cite{SUMO2018} generated network in the form of the average travel time for each sell to all others using the network. For an efficient computation with a large number of cells, we introduce the concept of neighborhoods by partitioning the cells according to their closest street section and reducing the problem to the computation of the shortest path from the target to source nodes of each of street represented as an edge in the network. 

Using the average travel time as a base measure of accessibility for all cells, the agent is then placed in the grid and tries to maximize its reward. We define the agent's current reward as the sum of the values of all developed sites, while choosing a structure type to be built belonging to one of the previously defined classes (residential, commercial, industrial and recreational). This choice becomes trickier as the construction of a given building may change the value of those previously developed. For example, adding an industry to a given region, while profitable on its own, could cause the value of neighboring residences to decrease. The same way that adding a few residences somewhere could increase the profitability of building a market nearby.
To represent that, we first define how the value is determined by each development type, mainly taking into account its neighboring cells and the calculated road network-aware accessibility. 

In the next section we discuss the algebraic manipulation done to find an efficient integration of the graph representing the road network with the rest of the map. Following that, we describe the value functions that govern the builder agent's choices as well as the training process.

\subsection{Road network integration}

As previously stated, one of the core features of the \textit{LUTI} models is the description of how the value, and therefore choice of locations in the urban environment, is greatly influenced by the accessibility, which in turn is mostly defined by the distances and availability of transport between each establishment.  

\begin{figure}[ht]
\centering
\includegraphics[width=.7\textwidth]{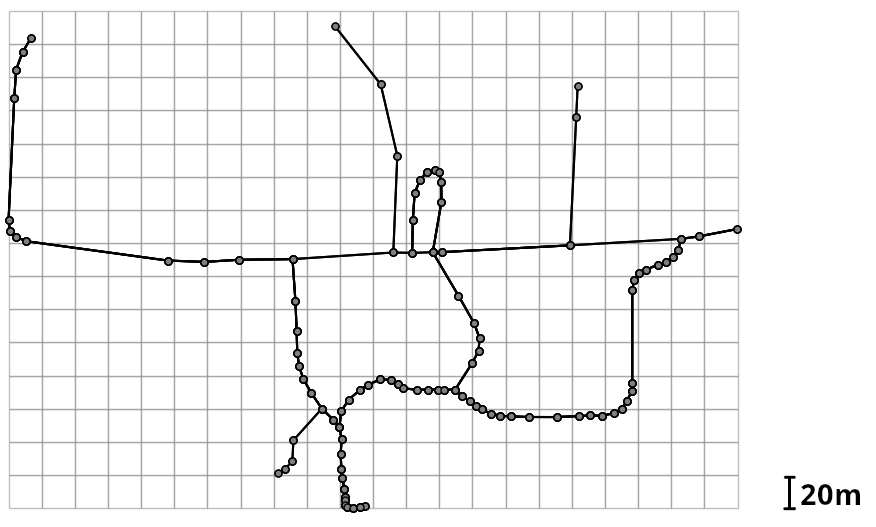}
\caption{Graph representation of a road network along with discretization of the terrain using a grid. Edges constitute street segments while vertices are changes in the geometry as well as any other property of that section. For the sake of readability, while our road networks are modeled as directed graphs, arrows representing the directions will be omitted in the illustrations.}
\label{fig:exampleFig1}
\end{figure}

The main challenge of efficiently modelling this interaction in an integrated system is the different structures most commonly used to represent each of the components. The road network, for example, is most naturally depicted as a graph where streets might typify edges and changes in their components such as speed limit, number of lanes or intersections are vertices, like illustrated in Figure \ref{fig:exampleFig1}. This representation also enables the use of a vast array of algorithms to effectively find paths and describe the overall structure of the network. For these reasons, we represent the road network as a directed weighted graph, where edges represent lanes and their respective directions.

As for the depiction of the terrain and land uses, a discrete, grid-like model greatly simplifies the dynamics of the system. This is specially true in an agent-based approach when defining interactions with such environment.

We solve this problem by treating those two elements as independent features of the model, coupled only by the shared coordinate system that enables the calculation of the distance from each cell to the network. That way, we can compute the precise travel time using the network from cells $c_i$ to $c_j$, where $i \neq j$, as:

\begin{equation}
    t(c_i, c_j) = t_N(c_i) + \frac{d(p_{c_i}, e_{t}^{c_i})}{v_{e^{c_i}}} + Dijkstra(e_{t}^{c_i}, e_{s}^{c_j}) + \frac{d(p_{c_j}, e_{s}^{c_j})}{v_{e^{c_j}}} + t_N(c_j)
\end{equation}

\begin{figure}[ht]
\centering
\includegraphics[width=.8\textwidth]{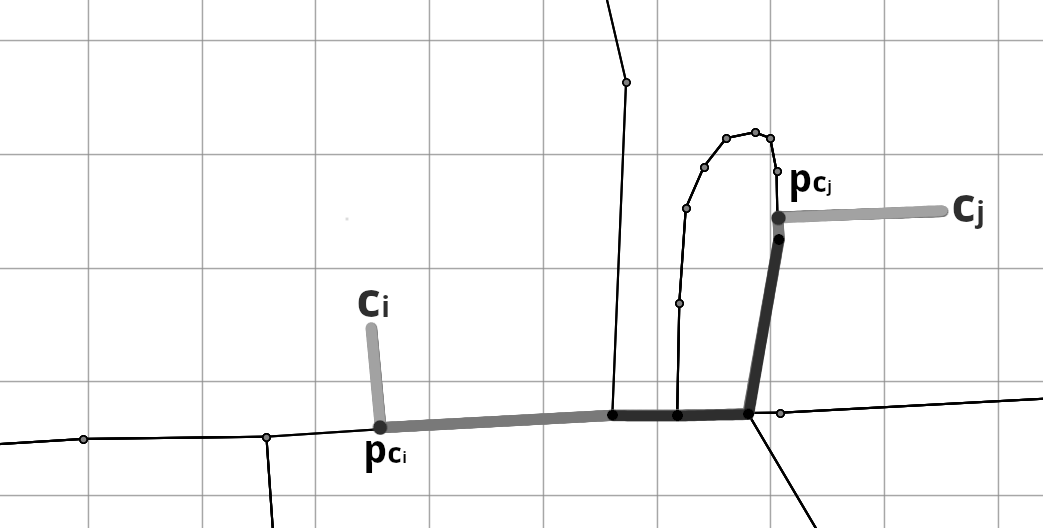}
\caption{Visual representation of the path taken from one cell to another. Starting from the origin cell $c_i$, first we compute the distance to the closest street. After that, from the point closest to $c_i$, we reach the target node with the speed of the edge and after that travel in the network using the shortest path to reach the source node of the edge that contains $p_{c_j}$.}
\label{fig:exampleFig3}
\end{figure}

Where for a given edge $e$, $v_e$ is the speed limit in the street segment represented by that edge. We start by defining $e^{c_i}$ and $e^{c_j}$ as the edges of the network $N$ closest to the cells $c_i$ and $c_j$ respectively. Taking an arbitrary edge $e$, we also define the vertices $e_s$ and $e_t$ as its source and target nodes. By doing so, we are able to split the travel between cells in four components like shown in Figure \ref{fig:exampleFig3}: the time spent going from the origin cell to the network, $t_N(c_i)$, then travelling from the closest point in the network to $c_i$, $p_{c_i}$, to the target node of the edge which $p_{c_i}$ lies on, $e^{c_i}_t$. After that, we use the length and speed limit of the street segments to define the weight of the network edges as the travel time and use the \textit{Dijkstra} algorithm to compute the shortest path between $e^{c_i}_t$ and $e^{c_j}_s$, the source node of the edge closest to the destination cell, $c_j$. Finally, just like the path from $c_i$ to $e^{c_i}_t$, the path form $e^{c_j}_s$ to $c_j$ is calculated and we obtain the travel time for the whole trip by summing all parts.

It is worth noting that by doing so, we assume that a path from $e^{c_i}_t$ to $e^{c_j}_s$ exists in the network. By restricting the roads to only connected networks, we assure this conditions is met for all pair of vertices by making sure it is strongly connected. We do so by checking if for every two nodes $n_i$, $n_j$, if $e=(e_s, e_t, w) \in E$ where $e_s = n_i$ and $e_t = n_j$, and $w$ is the weight of the edge exists in the network, there is also a corresponding edge with inverse direction $\epsilon = (n_j, n_i, \omega)$. In case $\epsilon \notin N$, a new edge $e_{inv} = (n_j, n_i, e_{len}/v_0)$ is added to the network. Doing that is equivalent to allowing movement in the opposite direction in a lane, while $e_{len}$ is the length of the street segment represented by edge $e$ and $v_0$ a speed defined by the user and usually much lower than that of $e$ as a punishment for walking in the wrong direction in the network and discouraging it anywhere other than where it is strictly necessary to reach the destination. In this work, we fix $v_0 = 5 km/h$. We also use this value when defining $t_N(c) = d_N(c)/v_0$, since by doing so the accessibility of cells further away from the network is also reduced. As result, we are able to obtain the average travel time from a given cell $c \in C$ to all others by using:

\begin{equation}
\label{eqn:sum1}
    T(c) = \frac{1}{|C|} \sum_i t(c, c_i)
\end{equation}

But the computational cost of doing so is considerably high, limiting the scale of the map we can represent. However, once we consider the sum over all cells, even thought the time travelling from one cell to another is usually different to that if we swap origin and destination, a lot of terms end up repeating. To be able to see that more clearly, we partition the set of all cells $C$ in the sets $\{C_{e_1}, C_{e_2}, ..., C_{|E|}\}$ where $E$ is the set of all edges and $C_e$ is the set of cells that has $e \in E$ as its closest edge.

Using the properties that $\cup_e C_e = C$ and $C_{e_i} \cap C_{e_j} = \O$ for $i \neq j$, we define the total time spent in the vicinity of a given edge $e$ as:

\begin{equation}
    T_e = \sum_{c \in C_e} \frac{d(p_c, e_s)}{v_e} + t_N(c)
\end{equation}

The advantage of this approach is that while the network is not changed, $T_e$ remains constant for all edges. The same is true for the shortest paths and distances in the network, such that for any cell $c$, the average travel to all others in a neighborhood $C^e$ different than its own (i.e. that do not share the closest edge) can be computed by taking:

\begin{equation}
    T_{C^e}(c) = t_N(c) + \frac{d(p_{c}, e_{t}^{c})}{v_{e^{c}}} + Dijkstra(e_{t}^{c}, e_{s}) + \frac{1}{|C_e|} T_e
\end{equation}

One more thing we should consider is travel within the same edge. While it is possible to just use the same approach that was described before, that would most likely imply that a whole travel through multiple streets would be needed to arrive at a neighbor cell, the most extreme case being that $t(c,c) \neq 0$. A few approaches can be used to tackle this problem, but it mostly consists of a trade-off between efficiency and validity of final result when compared to the actual average travel time. The most realistic option, producing the exact value of the average travel time can be attained by using the values of $T_{C^e}$ for all other neighborhoods while manually defining and calculating $t(c_i, c_j)$ if $c_i$ and $c_j$ share the closest edge. This gives us the average travel time:

\begin{equation}
    T(c) = \frac{1}{|C|} \left( \sum_{c_i \in C_{e^c}} t(c, c_i) + \sum_{e \in E, e \neq e^c} |C_e| T_{C^e}(c) \right)
\end{equation}

\begin{figure}[ht]
\centering
\includegraphics[width=.45\textwidth]{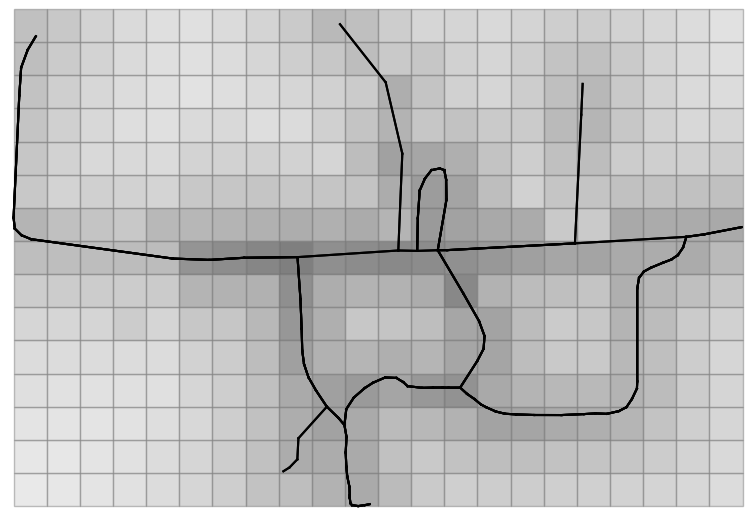}
\caption{
Map colored based on accessibility, darker 
cells corresponding to a lower average travel time in the map and, therefore,
higher accessibility.}
\label{fig:exampleFig4}
\end{figure}

As for the travel time within the same neighborhood, we define for $c_i, c_j$ that share the closest edge $e$, if $c_i \neq c_j$:

\begin{equation}
    t(c_i, c_j) = t_N(c_i) + \frac{d(p_{c_i}, p_{c_j})}{v_{e}} + t_N(c_j)
\end{equation}

while for $c_i=c_j$, $t(c_i, c_j) = 0$. The resulting map, with cells colored by accessibility is shown in Figure \ref{fig:exampleFig4}.

Note that while still scaling quadratically with $|C|$, effectively the sets $C_e$ are much smaller than $C$, which greatly reduces the computations. For the purposes of this paper, we will use this approach while handling maps of small to moderate size (up to $40000$ cells or an area of $16km^2$). 

After this first calculation is done for the whole map, the cells intersected by the road network are set as streets and become immutable during the land use classification process. The remaining cells are given an accessibility score that is the inverse of the average travel time and proceed to be explored by the agent.

\subsection{Score determination}

When the agent is initialized, it is given a list of available sites to  develop, determined by the cells located a maximum distance of $50m$ to the road network and ordered by their previously calculated accessibility scores. The agent then starts from the most accessible cell and moves down to the least accessible ones while in each step choosing between one of the development types to classify the undeveloped cell it is current standing on. Cells that are intersected by the road network, which is determined by their respective distance to the closest edge, are set as streets and marked as unavailable for development.

We set states as tensors of dimension $(2K+1) \times (2K+1) \times 6$, where $K$ is the number of cells to each side of the agents that it is able to observe. With the agent in the center, the information of the nearby cells is then represented as a set of six matrices: one for each development type (including the streets), composed of zeros or ones depending on whether the cell in that particular coordinate with respect to the agent is of said type, and a sixth matrix containing the normalized accessibilities.

Rewards are defined as the sum of the individual values of each cell. While undeveloped cells and streets have a value of $0$, the value of the developed ones are calculated based on how its respective type of development responds to its surroundings. Considering each cell observes the set of cells $K$ positions to each side of it, each type of development is influenced by its vicinity as described below.

\subsubsection{Residential developments}

The score of a residential development is defined as being positively related the total number of other residences surrounding it. The same is true for the number of commercial and recreational developments. On the other hand,  the presence of industries negatively impacts the total score.
We model these aspects considering the impact of the accessibility of the current location with the road network as:

\begin{equation}
    s_h = \tilde{a_c} \frac{W_{hh}Res(c, K) + W_{hc}Com(c, K) + W_{hi}Ind(c, K) + W_{hr}Rec(c, K)}{K^2}
\end{equation} 

where $\tilde{a_c}$ is the normalized accessibility of the cell, $Res(c, K), Com(c, K), Ind(c, K), Rec(c, K)$ the total number of residential, commercial, industrial and recreational developments in a radius of $K$ cells from $c$ and $W_{hh}$ $W_{hc}$, $W_{hi}$ and $W_{hr}$ the weight residential developments consider when evaluating the effects of those different types respectively.

\subsubsection{Industrial developments}

Different from residential developments, we model industrial developments in such a way that its value is only affected by other industries nearby. Moreover, instead of directly using the accessibility, the distance to the network is used directly, such that peripheral cells (further away from the center of the map) are preferred when trying to maximize the overall scores. With that, $s_i$ is given by:

\begin{equation}
    s_i = \exp{\left( -\frac{d_N(c)}{10} \right)} \frac{W_{ii}Ind(c, K)}{K^2}
\end{equation}

\subsubsection{Commercial developments}

Commercial developments are unaffected by industrial or recreational developments, while being sensitive to the total of residential developments and other commercial establishments due to effects on the prospective market. More specifically, while their scores are positively affected by the number of residences, other businesses are only valuable when not saturating the region's market. Thus, we define the score for commercial developments:

\begin{equation}
    s_c = \tilde{a_c} \left( \frac{W_{ch}Res(c, K) }{K^2} + W_{cc}\left( exp\left(\frac{- \left( Com(c, K) - Res(c, K) \right)^2}{K^2} \right) - exp\left( \frac{- Res(c, K)^2}{K^2} \right) \right) \right)
\end{equation}

Note that the middle term maxes out when $Ind(c, K) = Res(c, K)$. The rightmost term is used for the purpose of keeping the value within the interval $[1,0]$.

\subsubsection{Recreational developments}

As previously mentioned, the value of recreational developments is set as zero by default. Representing parks and other similar environments of the cities, their main purpose in our model is to increase the value of otherwise unwanted areas of the map. 

\subsection{Agent training}

We start by setting the weights for the scores. Trying to balance the scores of the different development types, we define the weights as shown in Table 1.

\begin{table}[ht]
\centering
\caption{Score weights for each development type.}
\label{tab:exTable1}
\begin{tabular}{lr}
\hline
\multicolumn{1}{|l|}{Weight}         & \multicolumn{1}{l|}{Value} \\ \hline
\multicolumn{1}{|l|}{$W_{hh}$} & \multicolumn{1}{r|}{1}     \\ \hline
\multicolumn{1}{|l|}{$W_{hc}$} & \multicolumn{1}{r|}{3}     \\ \hline
\multicolumn{1}{|l|}{$W_{hi}$} & \multicolumn{1}{r|}{-4}    \\ \hline
\multicolumn{1}{|l|}{$W_{hr}$} & \multicolumn{1}{r|}{6}     \\ \hline
\multicolumn{1}{|l|}{$W_{ch}$} & \multicolumn{1}{r|}{3}  
\\ \hline
\multicolumn{1}{|l|}{$W_{cc}$} & \multicolumn{1}{r|}{6}  
\\ \hline
\multicolumn{1}{|l|}{$W_{ii}$} & \multicolumn{1}{r|}{3}  
\\ \hline
\end{tabular}
\end{table}

The agent is placed in the cell with highest accessibility value and starts to randomly develop as it moves to the other cells while keeping track of the transitions $(s_{i}, a_{i}, R_{i+1}, s_{i+1})$ in its replay memory. For a given time step $i$, we define $s_i$ as the current state, $a_i$ as the action taken by the agent, $R_{i+1}$ the reward and $s_{i+1}$ the next state reached after taking action $a_i$ in state $s_i$. Rewards are calculated each time step by recalculating and summing the individual scores after the agent's action.

\begin{figure}[ht]
\centering
\includegraphics[width=.4\textwidth]{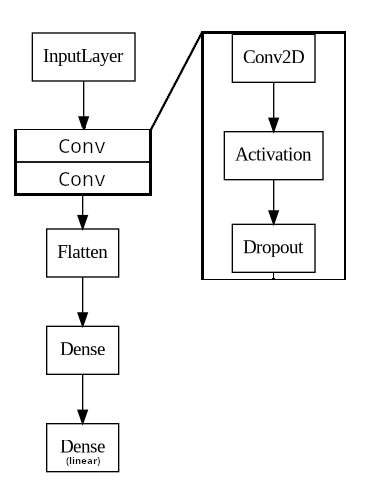}
\caption{
Neural network used for projecting the q-values. States are used as inputs in batches of 64 and than go through two sets of convolutional, activation and dropout layers. After being flattened, and serving as input to a dense layer, a final layer with linear activation produces the q-values for each of the four actions (one for building each type of land use).}
\label{fig:exampleFig5}
\end{figure}

Those transitions are used in the training of the neural network shown in Figure \ref{fig:exampleFig5} for predicting the q-values for state-action pairs. Taking the state tensors as inputs in batches of size 64 from the replay memory, those are fed into a convolutional neural network that predicts the q-values while being fitted against the actual attained rewards.

The training is done using two models: the base, that is fitted during each iteration, and the target, used to predict future q-values from the resulting states of the transitions. The target model is updated at the end of each episode, composed of either a set number of interactions of the agent with the environment or until the map runs out of undeveloped areas. 

We use maps gathered from random urban regions using \textit{OpenStreetMap} for training as the agent interacts with them and stores the experience in the replay memory. For the exploration/exploitation balance, we use an $\varepsilon$-greedy approach with $\varepsilon_0 = 1$ and decay rate of $0.999$ per episode. Interaction is done until the map is completely developed and each episode takes place in a new, randomly selected map.

\section{Evaluation}

\begin{figure}[ht]
\centering
\includegraphics[width=.7\textwidth]{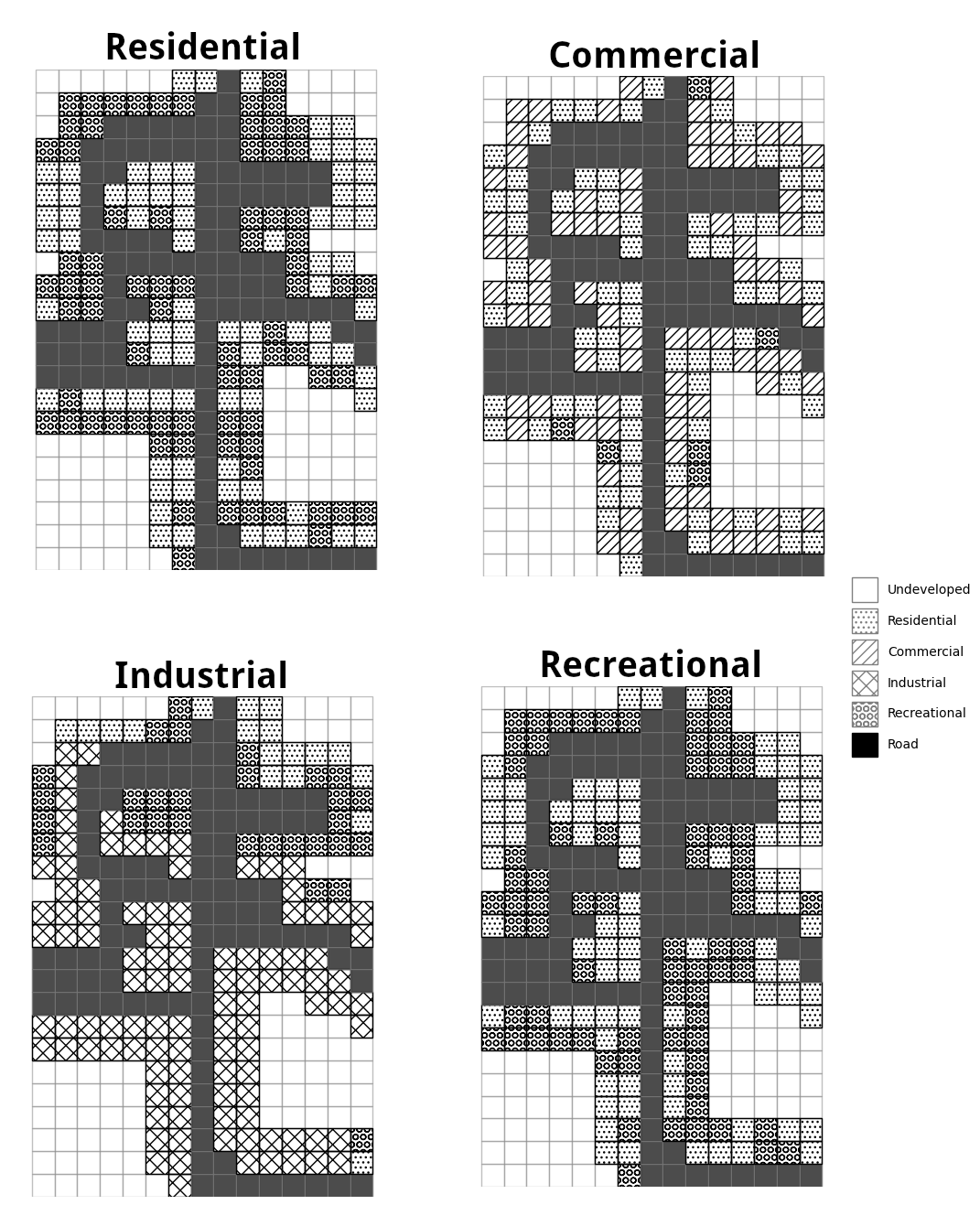}
\caption{
Resulting maps when taking different actions in the first step and then proceeding to maximize immediate reward.}
\label{fig:exampleFig6}
\end{figure}

We compare the performance of the trained agent with the average reward obtained from randomly developing the cells as well as choosing the action that maximizes the immediate reward. In our current model, the main challenge with the greedy approach is how we handle cases where the rewards gained from taking two different actions are the same. This is seen mainly when the agent builds in an empty neighborhood, since the reward remains unchanged. Moreover, the choice made in the first action, where aside from developing a commercial establishment the scores remains zero, can drastically change the evolution of the map, as seen in Figure \ref{fig:exampleFig6}.

\begin{figure}[!ht]
\centering
\includegraphics[width=.9\textwidth]{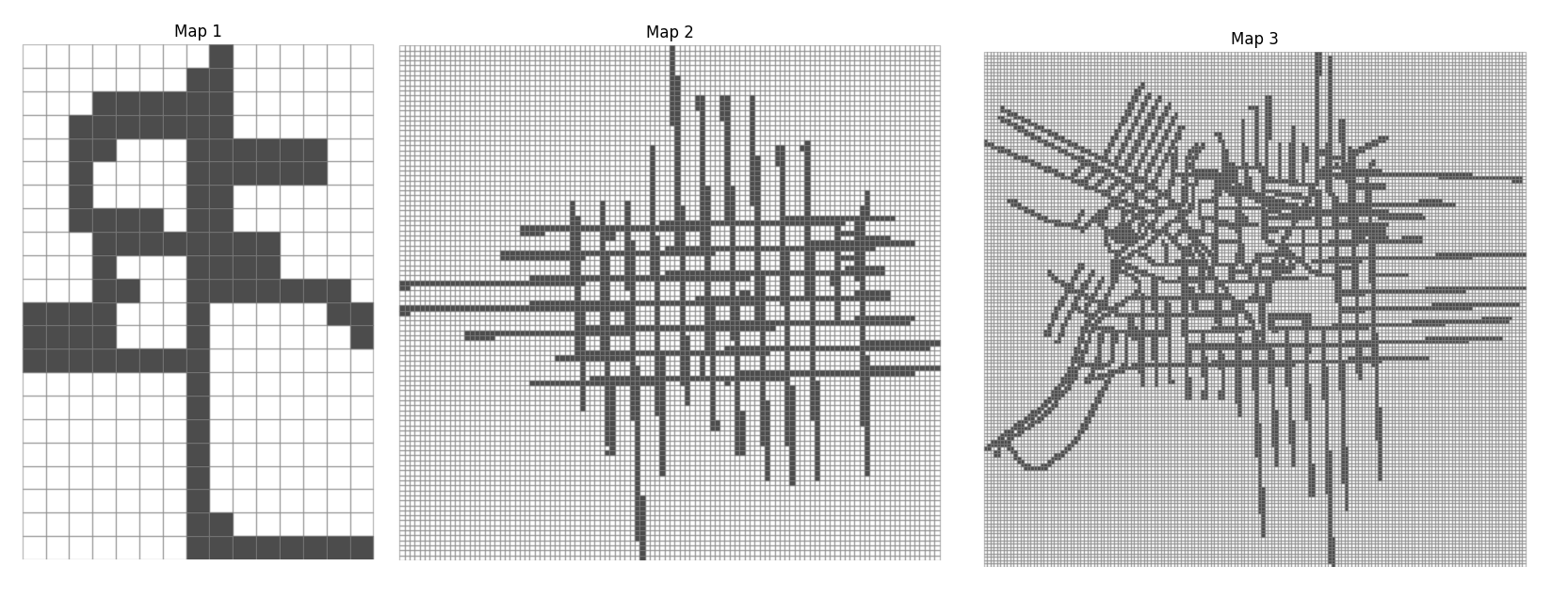}
\caption{Maps used for testing the agents applying different policies. Networks were extracted from real cities and used in the agent environment.}
\label{fig:exampleFig7}
\end{figure}

We handle this problem by randomly selecting the action in those cases and taking the average score over a thousand episodes. Results in three different maps are shown in Table 2, while the maps used for testing can be seen in Figure \ref{fig:exampleFig7}.

\begin{table}[ht]
\centering
\caption{Reward attained by agents after completely developing the maps. Here we note a clear advantage of the trained agent, seeking an optimal policy, over the greedy policy.}
\label{tab:exTable1}
\begin{tabular}{cccc}
\hline
\multicolumn{1}{|c|}{Agent policy} & 
\multicolumn{1}{|c|}{Map 1} &
\multicolumn{1}{|c|}{Map 2} &
\multicolumn{1}{|c|}{Map 3}\\ \hline\hline
\multicolumn{1}{|l|}{random} & \multicolumn{1}{|r|}{126.4} & \multicolumn{1}{|r|}{5122.2} & \multicolumn{1}{|r|}{10840.3}     \\ \hline
\multicolumn{1}{|l|}{greedy} & \multicolumn{1}{|r|}{212.5} & \multicolumn{1}{|r|}{14131.8} & \multicolumn{1}{|r|}{26870.2}     \\ \hline
\multicolumn{1}{|l|}{trained} & \multicolumn{1}{|r|}{\textbf{423.7}} & \multicolumn{1}{|r|}{\textbf{16190.1}} & \multicolumn{1}{|r|}{\textbf{30125.2}}     \\ \hline
\end{tabular}
\end{table}

\section{Conclusion}

As this work focuses mainly on the basic structure of the environment, once those fundamentals are established, new methods for modelling similar systems with the same goal can be more easily developed, expanding upon the current scenery of procedural generation of cities.

One of the core advantages of microsimulation and a bottom-up approach to modelling the urban scenery is the sheer breadth of characteristics and richness of details that can be incorporated in the systems \cite{perez}. As much as we try to take advantage of that in the form of the road network-grid integration, for example, much like in its case, computational bottlenecks are a recurring issue and the need for numerical and analytical methods for integrating those details while maintaining the scale of the simulation are needed.  

Similar applications that explore the robustness of \textit{LUTI} models tend to be more focused in one or only a few components that make up settlements. In our case, zoning takes the spotlight, however road network expansion is another highly studied field, combining both the geometric features and effects in accessibility to generate streets \cite{rui}.

For future works we plan to combine more of those aspects that constitute the bases of cities in a single system, maintaining the scale while building upon novel methods for city generation and expansion while incorporating real world data when modelling the environment as well as for evaluating the results.

\section{Acknowledgements}

The present work was financed by the Fundação de Amparo a Pesquisa do Estado de São Paulo (FAPESP) and supported by JSPS grant 22K11918.

\bibliographystyle{sbc}
\bibliography{sbc-template}

\end{document}